\documentclass[11pt]{article} 
\usepackage{hyperref} 
\usepackage{epsfig}
\pdfoutput=1

\begin{document} 
\title{Droplet breakup in homogeneous and isotropic turbulence}
\author{Federico Toschi$^{1}$, Prasad Perlekar$^{1}$, Luca
  Biferale$^{2}$, and Mauro Sbragaglia$^{2}$\\
  $^{1}$ Department of Physics,\\Department of Mathematics and Computer Science,\\
  and J.M. Burgerscentrum, Eindhoven University of Technology,\\ 5600
  MB Eindhoven, The Netherlands.\\
  $^{2}$  Department of Physics and INFN,\\
  University of Tor Vergata,\\Via della Ricerca Scientifica 1, 00133
  Rome, Italy.\\
  (International Collaboration for Turbulence Research)\\}
\maketitle
\begin{abstract} 
  This fluid dynamics video shows the breakup of a droplet in a
  stationary homogeneous and isotropic turbulent flow.  We consider
  droplets with the same density of the transporting fluid. The
  droplets and the fluid are numerically modelled by means of a
  multicomponent Lattice-Boltzmann method. The turbulent fluid is
  maintained through a large scale stirring force and  the radius of
  stable droplets, for the parameters in our simulation, is larger
  than the Kolmogorov scale.  Events of droplet deformation, break-up
  and aggregation are clearly visible from the movie. With the present
  database droplet evolution can be studied from both an Eulerian and
  Lagrangian point of view. The Kolmogorov-Hinze criteria for droplets
  break-up can be tested also by means of simulations with different
  viscosity contrast between the two components.
\end{abstract}

\section{Introduction}
Droplet emulsions are key to many natural and industrial processes. In
presence of an external flow, droplets undergo deformation, breakup,
and coagulation. In a turbulent flow, breakup of droplets larger than
the Kolmogorov scale is governed by the interplay between surface
tension and turbulent pressure fluctuations. Balancing surface tension
and pressure fluctuations allows to estimate the minimum unstable
droplet diameter that can undergoes breakup \cite{1,2}.  Turbulent
pressure-driven and surface-tension stresses across a droplet of
typical diameter $d$ and with surface tension $\sigma$ can be
estimated as:
\begin{figure}
  \includegraphics[width=.32\linewidth]{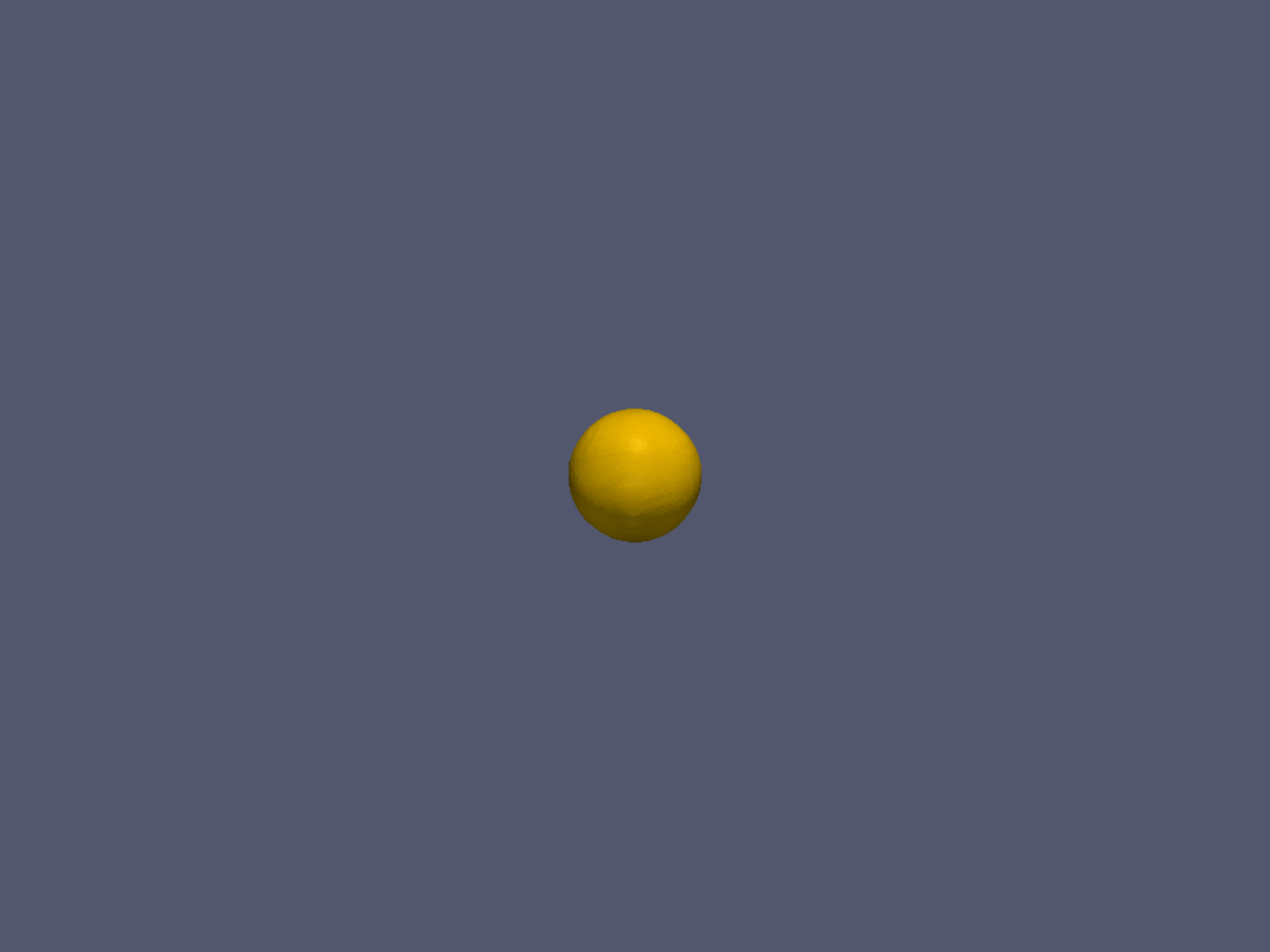}
  \includegraphics[width=.32\linewidth]{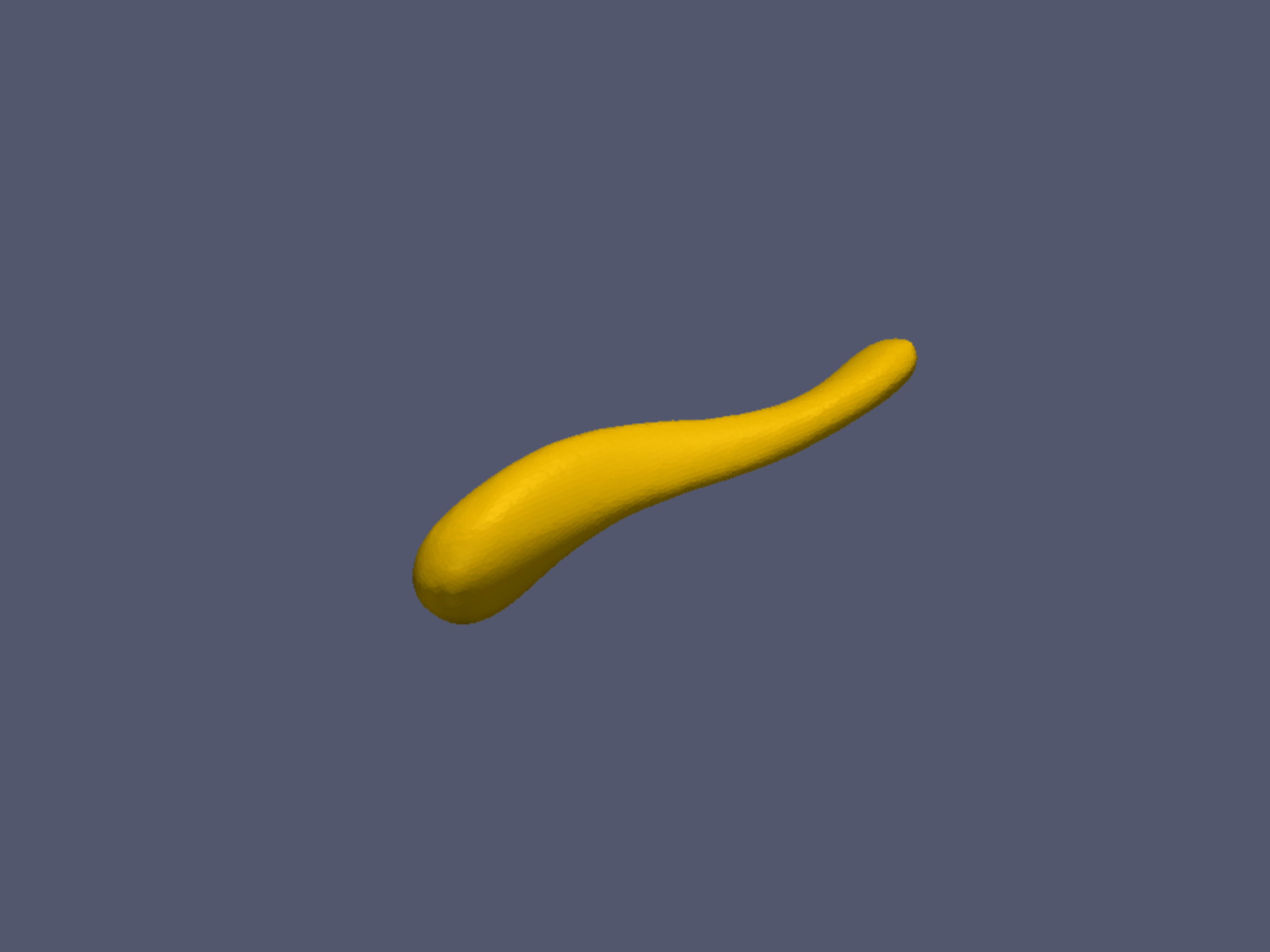}
  \includegraphics[width=.32\linewidth]{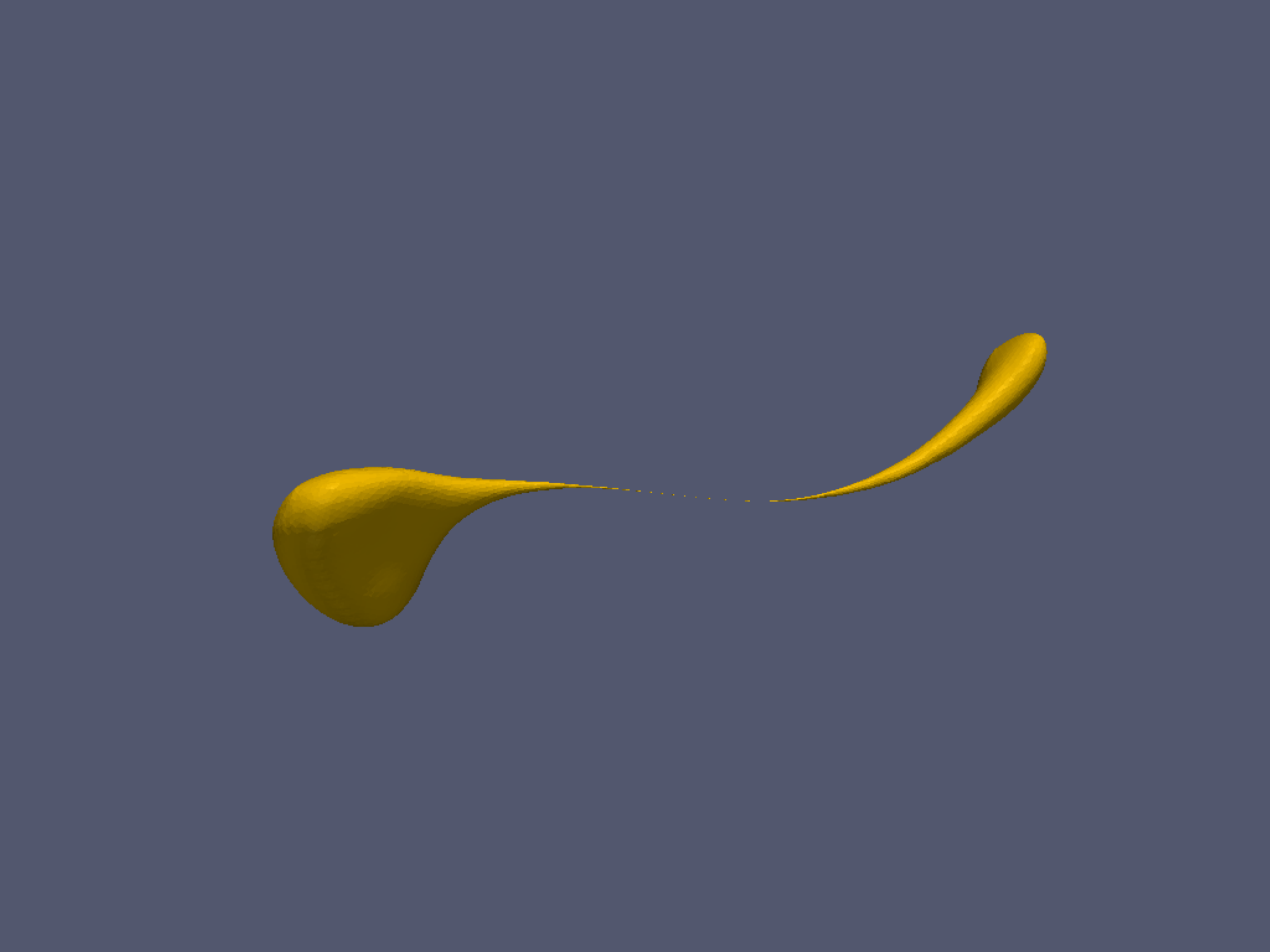}
  \caption{\label{gialla} The initially spherical droplet (left panel)
    get strechted (central panel) by switching on the stirring force
    and generation of a turbulent flow. The righ panel shows the
    initial droplet immediately after its first breakup.}
\end{figure}
\begin{figure}[!t]
  \includegraphics[width=\linewidth]{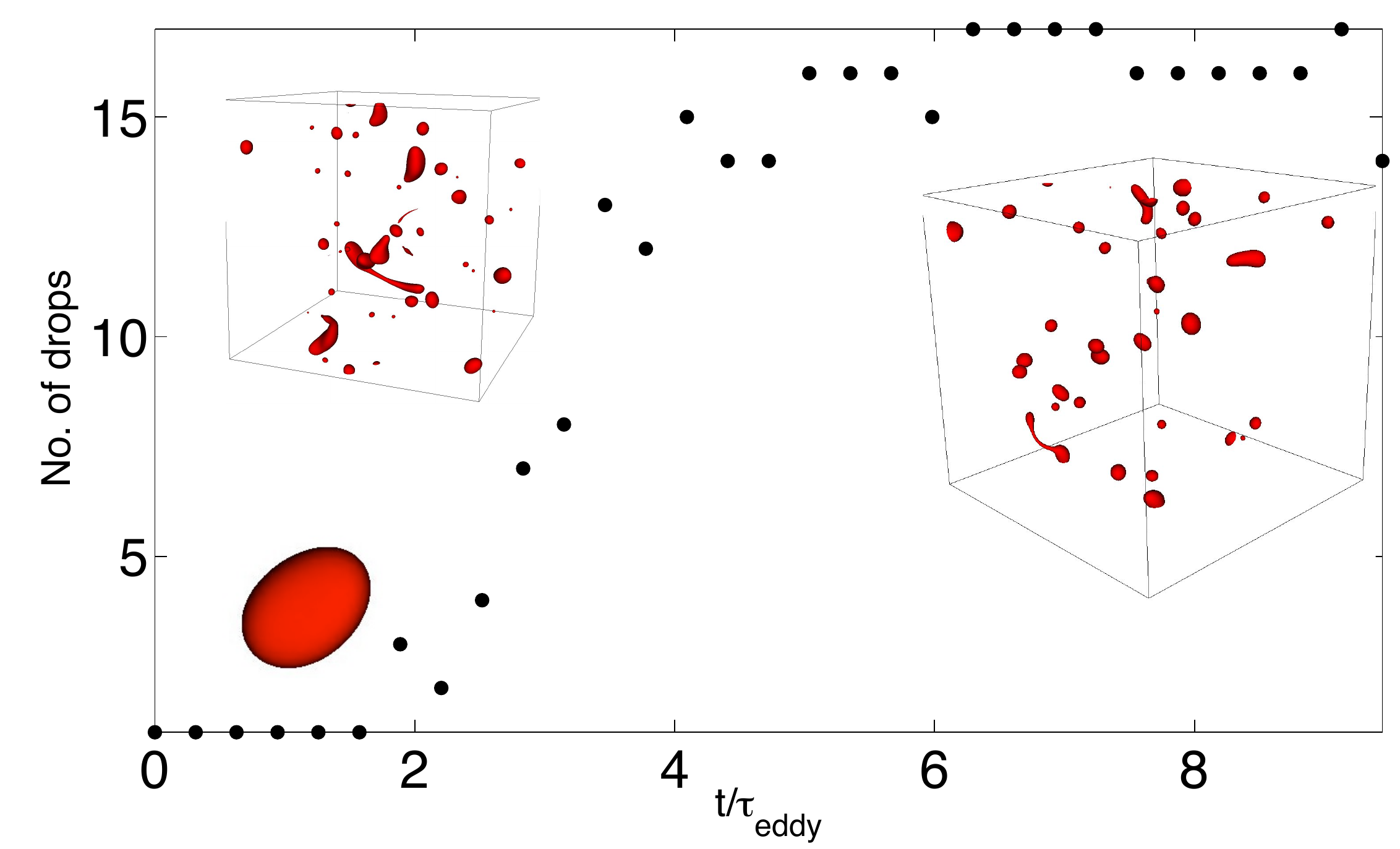}
  \caption{\label{phases} Different phases in the time evolution of
    a large droplet in a turbulent flow ($\bullet$). Initially the
    droplet deforms until it is broken it two droplets. This event is
    rapidly followed by several other multiple fragmentation events
    until a stationary droplet dispersion is attained. In the final
    stages the droplets continuously undergo deformation, break-up,
    and coagulation. Here the time is non dimensionalized by the large
    eddy turnover time $\tau_{\mbox{\tiny eddy}}$.}
\end{figure}
$$
\tau_{\mbox{\tiny turb}} \propto \rho_c \langle \left(\delta_d
  u\right)^2\rangle; \qquad \tau_{\mbox{\tiny{tens}}} \propto \sigma/d
$$
where with $\rho_c$ we indicate the density of the continous phase and
with $\langle (\delta_d u)^2\rangle$ we indicate a typical velocity
fluctuations across the droplet.  The ratio between the two gives the
Weber number estimated on the droplet diameter:
$$
We(d) = \frac{d \rho_c \langle (\delta_d u)^2\rangle}{\sigma}.
$$
The maximum stable {\it mean} droplet diameter can be estimated by
assuming K41 turbulent statistics for velocity increments: $\langle
(\delta_d u)^2\rangle \sim \varepsilon^{2/3} d^{2/3}$ where
$\varepsilon$ is the energy dissipation. From the condition of Weber
number order unity, one obtains:
\begin{equation}
  d_{c} \propto \left(\frac{\sigma}{\rho_c}\right)^{3/5} \varepsilon^{-2/5}.
\label{kh}
\end{equation}
Corrections to finite viscosity of the dispersed phase can also be
added \cite{3}.  The present simulation allows to study droplets
breakup conditions e.g. testing the Kolmogorov-Hinze criteria at
changing the Reynolds number and the viscosity ratio between the
dispersed and continuos phase. Specifically it is also possible to
study the fluctuactions around the mean average stable droplet radius
(\ref{kh}) induced by intermittent fluctuations of the velocity field.

In the movie various stages of droplet breakup process are shown. The
turbulent flow is homogeneous and isotropic and it is numerically
integrated by means of a multi-component Lattice-Boltzmann (MCLB)
simulation integrated in order to evolve consistently droplets and the
advecting fluid \cite{sundar}. The multicomponent flow is simulated
using a standard Shan-Chen model \cite{sc}, the turbulence is
generated by applying a large scale forcing, and periodic boundary
conditions are used.  The parameter used in our simulation are given
in table~\ref{table1}.

We start the simulation with the fluid at rest and all the droplet
volume fraction residing into a large, single spherical droplet. From
the beginning of simulation, the large scale stirring start to be
applied and the turbulence level grows continuosly during roughly one
large scale eddy turnover time. In the initial phase the large drop
gets mildly deformed, then stretched, and finally broken into smaller
droplets (see Figure \ref{gialla}). At this point the fragmentations
process continues for about another $1\div 2$ eddy turn over
times. After this period the distribution of droplets is almost
stationary and is characterized by continuos coagulation, deformation
and breakups events due to the continuous presence of underlying
turbulent fluctuactions. For very long simulation one may experience
some droplets evaporation. The different phases in the time evolution
of droplets are schematically illustrated in Fig.~\ref{phases}.

Interesting questions concern the rate of collision and the nature of
deformation and breakup events. It is also of interest to quantify the
evolution of the statistical properties of droplets diameter and the
correlations between stresses on the droplet and break-up
dynamics. These results will be reported in a forthcoming article
\cite{5}.

We acknowledge support from the Juelich supercomputing center (Germany) for the computational resources.

\begin{table}
\begin{center}
\begin{tabular}{@{\extracolsep{\fill}} c c c c c c c c c}
  \hline
  &$N$ & $\sigma$ & $\nu$ & $u_{\mbox{\tiny rms}}$ & $R_{\lambda}$ & $d_0$ & $d_c$ & $d_c^{lbm}$  \\
  \hline \hline
  & $512$ & $1.6\times 10^{-3}$ &$5\times10^{-3}$& $8.5\times10^{-3}$ & $29$ & $50$ & $24.16$ & $25\pm 2$ \\
  \hline
\end{tabular}
\end{center} \caption{\label{table1} Parameters of the numerical
  simulation from which the movie was produced. $N$ is the number of
  grid points along each direction, $\sigma$ is the surface tension,
  $R_{\lambda}\equiv u_{\mbox{\tiny rms}}\lambda/\nu$ is the
  Taylor-microscale Reynolds number, $\lambda \equiv \sqrt{E/\Omega}$
  is the Taylor-microscale, $E$ is the average fluid kinetic energy
  and $\Omega$ is the average enstrophy.  $d_0$ is the initial droplet
  diameter, $d_c$ is the critical droplet diameter obtained from the
  theoretical estimate (see text) and $d_c^{LBM}$ is the critical
  droplet diameter as measured from our simulation.  The Reynolds
  number was kept low on purpose in order to keep the dissipative
  scale relatively large: this allow us to have an interface thickness
  smaller than the viscous scale.}
\end{table}

\end{document}